\documentclass[a4paper,12pt]{article}
\usepackage{INTERSPEECH2016}
\usepackage{graphicx}
\usepackage{amssymb,amsmath,bm}
\usepackage{textcomp}
\usepackage{graphicx}
\usepackage{wrapfig}
\usepackage{lipsum}
\usepackage{dblfloatfix}
\usepackage{fixltx2e}
\usepackage{nameref}
\usepackage{amsmath}
\usepackage{hyperref}
\usepackage{graphicx}
\usepackage{varioref}
\usepackage{booktabs}  % nice looking tables
\usepackage{siunitx}
\usepackage{numprint}
\usepackage{textcomp}
\usepackage{color}

\usepackage{booktabs}  % nice looking tables
\usepackage{lipsum}

\title{A Speaker Diarization System for Studying Peer-Led Team Learning Groups}
\makeatletter
\def\name#1{\gdef\@name{#1\\}}
\makeatother \name{{\em Harishchandra Dubey, Lakshmish Kaushik, Abhijeet Sangwan, John H. L. Hansen\textsuperscript{+}\thanks{\textsuperscript{+}This project was funded in part by AFRL under contract FA8750-15-1-0205 and partially by the University of Texas at Dallas from the Distinguished University Chair in Telecommunications Engineering held by J. H. L. Hansen.}}}
\address{Center for Robust Speech Systems, Eric Jonsson School of Engineering\\
The University of Texas at Dallas, Richardson, TX 075080, USA \\
{\small \tt \{harishchandra.dubey, abhijeet.sangwan, john.hansen\}@utdallas.edu}
}
\begin{document}
\maketitle
% * <abhijeet.sangwan@gmail.com> 2016-03-30T12:03:35.635Z:
% \blfootnote{\textsuperscript{+}This project was funded in part by AFRL under contract FA8750-15-1-0205 and partially by the University of Texas at Dallas from the Distinguished University Chair in Telecommunications Engineering held by J. H. L. Hansen.}
% \footnote{\textsuperscript{+}This project was funded by AFRL under contract FA8750-15-1-0205 and partially by the University of Texas at Dallas from the Distinguished University Chair in Telecommunications Engineering held by J. H. L. Hansen.}
%-5mm}
\begin{abstract}
%-2mm}
Peer-led team learning (PLTL) is a model for teaching STEM courses where small student groups meet periodically to collaboratively discuss coursework. Automatic analysis of PLTL sessions would help education researchers to get insight into how learning outcomes are impacted by individual participation, group behavior, team dynamics, \emph{etc.}. Towards this, speech and language technology can help, and speaker diarization technology will lay the foundation for analysis. In this study, a new corpus is established called CRSS-PLTL, that contains speech data from 5 PLTL teams over a semester (10 sessions per team with 5-to-8 participants in each team). In CRSS-PLTL, every participant wears a LENA device (portable audio recorder) that provides multiple audio recordings of the event. Our proposed solution is unsupervised and contains a new online speaker change detection algorithm, termed $G^3$ algorithm in conjunction with Hausdorff-distance based clustering to provide improved detection accuracy. Additionally, we also exploit cross channel information to refine our diarization hypothesis. The proposed system provides good improvements in diarization error rate (DER) over the baseline LIUM system. We also present higher level analysis such as the number of conversational turns taken in a session, and speaking-time duration (participation) for each speaker. 
% Based on these experiments, the speaker diarization methodology proposed communication is shown to be effective for the study of small groups. 
\end{abstract}
\section{Introduction}
\footnote{\textcolor{blue}{This material is presented to ensure timely dissemination of scholarly and technical work. Copyright and all rights therein are retained by the authors or by the respective copyright holders. The original citation of this paper is:
Harishchandra Dubey, Lakshmish Kaushik, Abhijeet Sangwan, John H. L. Hansen, "A Speaker Diarization System for Studying Peer-Led Team Learning Groups", In the Proceedings of INTERSPEECH 2016, San Francisco, USA.}}
%\begin{figure*}[!t]
%\centering
%\includegraphics[width=500bp]{./figs/onlinediar.png}
%\caption{Naturalistic Audio Processing for simultaneously collected multi-stream audio data from Peer-led Team Learning (PLTL).}
%\label{fig_napc}
%\end{figure*}
%-2mm}
Peer-led team learning (PLTL) is a strategy used for improving learning outcomes in group settings for STEM students. Each team is led by a student who has already completed the course and is familiar with the course learning goals and/or challenges. The team lead coordinates the discussion on solutions of a given set of questions in study sessions. There is typically weekly study sessions held throughout the semester. PLTL is a popular approach and has been adopted by various universities at the undergraduate level. Additionally, education researchers have studied various aspects of PLTL to understand its impact on student's knowledge measured in terms of their success in academic programs~\cite{wamser2006peer,lyle2003statistical}. Typically, such research studies use control groups (by comparing students who do and do not participate in PLTL) and outcome metrics such as course grades or potentially opinions surveys to understand the educational impact. Here, analyzing actual student-to-student voice interaction in study sessions can help develop a richer understanding of how student success is related to participation, engagement, group behavior, team lead, benefits~\emph{etc.}. However, this would require analyzing large quantities of data and the use of speech and language processing tools would be especially beneficial. 

In this study, we explore the utility of speaker diarization technology in measuring simple communication metrics for PLTL sessions. Specifically, we describe a new corpus called CRSS-PLTL that was developed to facilitate this study. In CRSS-PLTL, we collected longitudinal data from 5 PLTL teams for one semester. Every PLTL session lasted for about 80 minutes where each team member wore a LENA audio recording device. Hence, the corpus contains multi-channel audio data for all sessions. This is different from typical diarization research that focuses on data collected using a single or multiple fixed far-field microphones ~\cite{vijayasenan2012multistream,vijayasenan2012diartk,gallardo2006multi}. It is common for students to physically move during PLTL sessions(\emph{e.g.}, walking to whiteboard to solve problems) as well as breaking-up into smaller groups for discussion. The speaking style is spontaneous and casual. Short conversation turns and overlapped speech are often encountered. All these factors make speaker diarization challenging for these scenarios.

Speaker diarization systems have been extensively researched, often for specific tasks ~\cite{miro2012speaker,tranter2006overview,yella2015comparison,ghaemmaghami2015cluster}. Both supervised and unsupervised methods have been explored. Quiet recently, some researchers have suggested a method for speaker diarization using Restricted Boltzmann Machines~\cite{pikrakis2014unsupervised}. The unsupervised methods for classification and segmentation of audio data has attracted attention in recent years~\cite{huang2006advances}. Among multi-stream diarization, meeting recordings have been analyzed by combining MFCC and TDOA features with various segmentation and clustering algorithms~\cite{vijayasenan2012multistream,vijayasenan2012diartk,gallardo2006multi}. In this study, we propose an unsupervised system for diarization suitable for studying PLTL groups. Particularly, we propose new unsupervised methods for speaker change detection and speaker clustering. In our experiments, we compared the proposed method with the LIUM diarization system. The proposed method in this study achieves more than 10\% absolute reduction in diarization error rate (DER) over LIUM for CRSS-PLTL data. Finally, we also use the diarization information to compute downstream metrics such as the number of conversational-turns taken and participation, and discuss how such metrics can assist in automatic analysis of PLTL groups.
\begin{figure*}[!t]
\centering
\includegraphics[width=450pt]{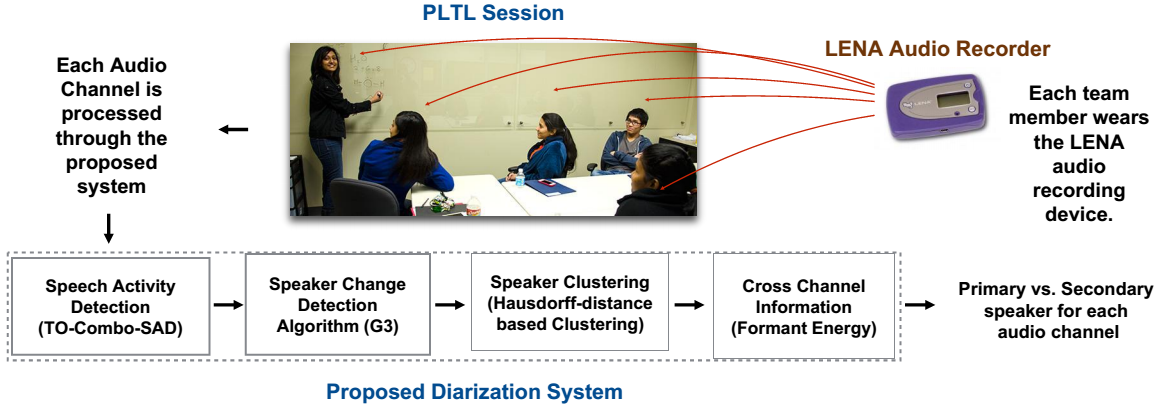}
%-1mm}
\caption{{\it In CRSS-PLTL, LENA audio recorders were worn by each team member for the entire session, that yields multi-channel audio data. The proposed speaker diarization system uses TO-Combo-SAD~\cite{ziaei2014speech} to remove non-speech segments, and then uses Unsupervised $G^{3}$ algorithm along with Hausdorff-distance based clustering to perform speaker change detection and clustering, respectively.}}
\label{fig_pltl}
%-2mm}
\end{figure*}
%%
%-2mm}
\section{Proposed System}
%-1mm}
Fig.~\ref{fig_pltl} shows the proposed system. As shown in the figure, each PLTL team member wears a LENA audio recorder unit (that essentially acts as a close-talk microphone). Therefore, each session yields multi-channel audio data where the number of channels was equal to the number of participants. This makes CRSS-PLTL corpus somewhat different from corpora typically used for diarization research where fixed far-field microphones are used for audio capture. This difference allows us to solve the overall diarization problem by solving primary speaker (person wearing the LENA device)~\emph{vs.} secondary speaker (all other speakers) detection problem for each audio stream. In other words, we were always solving a two-speaker diarization problem for every channel (we were interested in detecting the primary speaker, and categorize all other speakers as secondary). The overall diarization information can now be generated by merely combining primary speaker hypothesis from each audio channel. As seen in Fig.~\ref{fig_pltl}, Speech Activity Detection (SAD) is first performed to separate non-speech from speech. In this study, we used Threshold Optimized Combo-SAD (TO-Combo-SAD) for  SAD ~\cite{tocombosad,ziaei2014speech}. In the next step, the speech data is processed by the unsupervised $G^{3}$ algorithm that detects speaker change points and provides this information to the Hausdorff distance-based clustering algorithm that finds primary and secondary clusters. In what follows, we describe these algorithms in greater detail.
%The role of culture, gender and personality, intimacy and psychology in  a learning environment can be inferred by processing the simultaneously collected multi-stream audio from LENA devices worn by each student. These audio recording are also useful for assessing a student's comfort in a group study such as PLTL.  
%%
%-2mm}
\subsection{Unsupervised $G^{3}$ Algorithm}
%-1mm}
We propose a new method for unsupervised speaker change detection based on the work discussed in~\cite{ajmera2004robust}. Using the theoretical foundation provided in ~\cite{ajmera2004robust}, we investigated a large number of features and feature processing steps, and found a method that works well in practice for our data. We first extracted Mel-frequency cepstral coefficients(MFCCs) along with delta and delta-delta features (39-dimensions). The features were extracted for 40ms speech frames with 10ms skip rate. Additionally, a 320-dimensional real cepstrum of the linear prediction residual (RCLPR) is also used, since it models speaker-specific excitation information~\cite{prasanna2006extraction}. The 320 RCLPR features are then transformed with 51-point 1-D discrete cosine transform (DCT) to decorrelate the feature subset. Finally, the MFCCs and RCLPRs features are fused to form the final 90-dimensional fusion feature, that was used for speaker change detection.

Now, we describe the algorithm for speaker change detection. Let $\mathbf{U}$ and $\mathbf{V}$ be sets of fusion feature vectors taken from two successive 1-second time segments around time $'t'$ (we chose 1-second time window because we were interested in detecting short conversation turns, but this value can be adjusted as per application). Let $\mathbf{W}= [\mathbf{U}, \mathbf{V}] $ be the feature vectors of both frames. For detecting speaker change, we develop a binary hypothesis test, $H_{0}$ \emph{vs.} $H_{1}$, where $H_{0}$ denotes no speaker change at time $'t'$, and $H_{1}$ denotes speaker change at time $'t'$. To facilitate the test, we build models for both hypotheses. On one hand, we use a 2-component GMM (Gaussian Mixture Model) to model $\mathbf{W}$. On the other hand, we use simple Gaussian function to model $\mathbf{U}$ and $\mathbf{V}$ independently. Since, one \textbf{G}MM and two \textbf{G}aussians are used in this method, we name it $G^{3}$ algorithm. The GMM parameters are estimated on-the-fly using the expectation maximization (EM) algorithm. 

Now, let $\phi_{w}$ be the parameter vector of a 2-GMM estimated from $\mathbf{W}$, and let $\phi_{u}$ and $\phi_{v}$ be the Gaussian parameters for $\mathbf{U}$ and $\mathbf{V}$, respectively. If we assume the features in $\mathbf{U}$ and $\mathbf{V}$ are independent and identically distributed, we have the following expression for log likelihood $L_{H_{0}}$ and $L_{H_{1}}$ for both hypotheses $H_{0}$ and $H_{1}$, respectively,
%-1mm}
%
\begin{equation}
L_{H_{0}} = \sum_{i=1}^{N} \log(p(u_{i}|\phi_{w})) + \sum_{j=1}^{N} \log(p(v_{i}|\phi_{w})) ,
\end{equation}
and
%0mm}
\begin{equation}
L_{H_{1}} = \sum_{i=1}^{N} \log(p(u_{i}|\phi_{u})) + \sum_{j=1}^{N} \log(p(v_{i}|\phi_{v})) ,
\end{equation}
where $p(x|\phi)$ is the likelihood of the fused feature vector $x$ given model parameters $\phi$. The detection index, $D_{LLR}$, is based on log-likelihood ratio (LLR) and is given by 
%
%0mm}
\begin{equation}
D_{LLR} = L_{H_{1}} - L_{H_{0}} ,
\label{eqn_llr}
\end{equation}
%
%0mm}
where $D_{LLR}$ is greater than 0 whenever the 2-component GMM is a better model for the observed fused feature vector $\mathbf{W}$. Hence, speaker change ($H_1$) occurs when $D_{LLR} > 0$~\cite{ajmera2004robust}.
%
%-2mm}
\subsection{Hausdorff distance-based Speaker Clustering}
%-1mm}
Most state-of-the-art diarization systems used for TV shows and meetings tend to use hierarchical clustering. However, research has shown that spectral clustering that involves eigen-decomposition and k-means clustering is computationally simple as compared to hierarchical clustering~\cite{ning2006spectral}. For example, in~\cite{ning2006spectral}, the authors  used Japanese Parliament audio data that had segments of length 3 seconds or greater to compare hierarchical and spectral clustering. Spectral clustering is a global approach and hence optimal with respect to similarity criterion. On the other hand, hierarchical clustering is greedy and can lead to sub-optimal solutions. However, the performance of spectral clustering largely depends on the choice of similarity metrics. Here, Kullback-Leibler (KL) divergence is not the best suited for audio segments of less than 3 seconds~\cite{ning2006spectral}. In CRSS-PLTL, short speaker turns (about 1 second) were quite common, that made it difficult to use the KL divergence metric. This motivated the need to research a more suitable metric. In this study, we propose to use Hausdorff distance as similarity measure for spectral clustering. 

The Hausdorff distance assigns a scalar metric for similarity between two vectors or two matrices or a vector and matrix of different sizes. It has been found to be effective in tracking similarity among complex structures ~\cite{huttenlocher1993comparing,hclus}. Let $\mathbf{A_{1}}$ and $\mathbf{A_{2}}$ be feature matrices of dimension $m_{1} \times n$ and $m_{2} \times n$ where $m_{1}$ and $m_{2}$ are number of frames in both audio segments and $n$ being the feature dimension. The Hausdorff distance between feature matrices, $\mathbf{A_{1}}$ and $\mathbf{A_{2}}$ is given as 
%-1mm}
\begin{equation}
d_{H}(\mathbf{A_{1}},\mathbf{A_{2}})= \max (h(\mathbf{A_{1}},\mathbf{A_{2}}), h(\mathbf{A_{2}},\mathbf{A_{1}})) , 
\end{equation}
where $h(\mathbf{A_{1}},\mathbf{A_{2}})$ is given by,
%-1mm}
\begin{equation}
h(\mathbf{A_{1}},\mathbf{A_{2}})= \max \limits_{a_{1}\in \mathbf{A_{1}}} \min \limits_{a_{2}\in \mathbf{A_{2}}}\parallel a_{1} -a_{2}  \parallel        ,
\end{equation}
%-1mm}
and $\parallel \cdot \parallel$ is some underlying norm such as $L_{2}$ or Euclidean norm on elements in $\mathbf{A_{1}}$ and $\mathbf{A_{2}}$. Here, $d_{H}$ is the Hausdorff distance between two feature matrices, $\mathbf{A_{1}}$ and $\mathbf{A_{2}}$. Using Hausdorff distance as a similarity metric, various audio segments are compared and the most similar are merged together. Next, the Hausdorff distance between newly merged cluster and other clusters is recomputed and the process is repeated until we are only left with two clusters (one each for primary and secondary speakers).
%For separation of primary and secondary speakers in a PLTL channel, the number of clusters in 
%The LENA device is attached to the primary speaker and hence the audio signal from that channel in general comes from the primary speaker.
%\subsection{Spectral Clustering}
%-2mm}
\subsection{Primary speaker identification}
\label{sec:primaryspeakerid}
%-1mm}
Once two clusters are identified using Hausdorff distance based clustering, primary and secondary clusters are identified in the last step. The identification can be made based on a simple observation that the primary speaker tends to be closer to the microphone than secondary speakers. This causes primary speech to be more energetic than secondary speech. We have previously exploited this fact in other studies~\cite{ziaei2015prof,ziaei2014speech}, and have seen that this is a fairly robust assumption that tends to get even stronger with increasing duration. By measuring the average energy in the two clusters, we assign the cluster with higher and lower energies to primary and secondary speakers, respectively. The energy computation is performed by summing the energy of the first two speech formants. 

Finally, since we have multi-channel data, the energy measurements across channels can be further exploited to improve primary speaker identification. It is useful to note that while all microphones pick up every speaker's voice (due to close proximity), each speaker is loudest (most energetic) on their own microphone (owing to the physical distance separating speakers from the microphone). Additionally, it is assumed that overlapped speech is rare, and only one speaker speaks at a time (our analysis of the data showed that less than 3\% of the data contained overlapped speech). In other words, there is only one primary speaker across all channels at any given time. To exploit this, we scan decisions across all channels for fixed time windows (we used 2 second windows in our experiments), and identify regions where more than one channel contains the primary speaker. For these regions, we retain primary speaker decision only for the most energetic channel, and reverse the decision to secondary speakers for other channels. This process allows us to further refine the diarization hypothesis. There were some temporal shifts in various audio streams that was not utilized in this paper. 
%-2mm}
\subsection{Analysis}
%-1mm}
Once primary \emph{vs.} secondary speaker decisions are available for each audio channel, the overall diarization information is readily made by merely combining the individual channel results. Using the basic diarization information, a number of interesting metrics can be derived for the PLTL session. In this study, we show two metrics: (i) speaker turn-taking, and (ii) speaker participation measured using speech duration.

The quality of a conversation either in a classroom scenario such as PLTL or those at workplaces can be quantified qualitatively in terms of turn-taking. More turn-taking between various speakers in a group discussions shows more engagement and hence healthy discussions. For PLTL scenario, better engagement in solving tutorial problems can conclude that students are motivated in problem solving. We used the $G^{3}$  algorithm for counting the conversational-turns taken. Total number of conversational turns taken is given by the total number of speaker-changes for each channel of PLTL. Averaging the total-turns from each channel, we get the average turns taken in PLTL session. This metric quantifies the quality of discussions in that session. We compute the speaker-changes on a sliding segment of 1 second duration. The total conversational-turns computed from various channel are summarized in Table~\ref{table_turns}.
%
%-2mm}
\section{Experiments}
%-1mm}
\subsection{CRSS-PLTL Corpus}
%-1mm}
While collecting CRSS-PLTL corpus data, 5 PLTL teams were  tracked over an entire semester. Each team consisted of 5-to-8 members, where one member was always the team leader. All teams met once every week for a total of 11 weeks, resulting in a total of 55 sessions for the corpus. All students were part of an undergraduate Chemistry course. The collection is longitudinal as it tracks individuals over a 3-month time period. Each session was 80 minutes long, and each team member wore a numbered LENA audio recording unit for the entire duration of the session. It is useful to note that the LENA digital language processor (DLP) can record audio signals for long duration upto 16 hours and has been used for a variety of human-to-human communication research studies, especially adult-child interaction~\cite{xu2008signal,ziaei2013prof,sangwan2015studying}.The audio data in CRSS-PLTL contains varying amounts of noise and reverberation, and at times, the noise and reverberation level can be significantly degrading. Finally, each student completed a survey after each session that sought Likert-scale ratings for subjective questions such as behavior, communication, learning,~\emph{etc.}. In order to facilitate experimental evaluation for this study, 21 minutes from one session was chosen, and manual annotations for speech activity and diarization were created. This evaluation set contained 7 parallel audio channels (corresponding to 7 team members who attended that session). We downsampled the audio data to 8 kHz before processing it. It is same for all results discussed in this paper.
%-2mm}
%\begin{figure}[!t]
%\centering
%\includegraphics[width=250bp]{./figs/online_Diarization5.png}
%\caption{{ \it Primary and Secondary Speaker Diarization.}}
%\label{fig_napc}
%\end{figure}
%-2mm}
\subsection{Baseline System}
%-1mm}
  %Content Analysis for Audio Classification and Segmentation
 % Perceptual features are always good
We used the LIUM speaker diarization system as the baseline diarization system and compare its performance with the proposed system~\cite{gallardo2006multi,meignier2010lium}. The standard LIUM system was used for results presented in this paper. It is possible to use reasonable amount of labeled PLTL data for optimizing the LIUM system parameters. However, we have not optimized LIUM system for results discussed in this paper due to unavailability of enough labeled data. For all the experiments, the audio signals were downsampled at 8 kHz. The speech signal was divided into frames of size 40ms with a skip rate of 10ms. Our previous study has shown that TO-Combo-SAD worked better than the default SAD setup in LIUM~\cite{ziaei2014speech}. Hence, we used TO-Combo-SAD to generate speech \emph{vs.} non-speech decisions. We constrained LIUM to 2-speaker decisions, and further used the primary speaker identification method described in Sec.~\ref{sec:primaryspeakerid} to make primary \emph{vs.} secondary speaker decisions.
%-1mm}
\subsection{Results \& Discussions}
%-1mm}
We used DER as the figure of merit for the proposed and baseline diarization systems. DER, as defined by the NIST Rich Transcription Evaluation~\cite{nistder}, can be computed as, 
%-1mm}
\begin{equation}
DER = \frac{L_{fa} + L_{miss} + L_{err} }{L_{total}}
\label{eqn_der}
\end{equation}
%-1mm}
where where $L_{fa}$ is the total number of non-speech segments detected as speech, $L_{miss}$ is the total number of the speech
segments detected as non-speech, $L_{err}$ is the
total number of speech segments that were detected as speech but clustered as incorrect speakers, and $L_{total}$ is the total number of speech segments obtained using the ground-truth labels. Average DER across various channels was used as a metric for performance comparison. Additionally, we also compute and report equal error rate (EER) for TO-Combo-SAD. Table~\ref{table_baseline} shows DER and EER numbers for the baseline and proposed systems. Systems A, B and C are variations of the baseline LIUM system, where A is the LIUM system, B is LIUM system that takes SAD decisions from TO-Combo-SAD, C is LIUM system with TO-Combo-SAD that uses primary speaker identification described in Sec.~\ref{sec:primaryspeakerid}. As seen in the table, TO-Combo-SAD (8.67\% EER) delivers superior SAD decisions \emph{vs.} LIUM SAD (12.54\% EER). Furthermore, using TO-Combo-SAD and primary speaker identification reduces overall DER for the task by about 3\% absolute (35.80\% to 32.76\%). However, the proposed diarization system is able to significantly outperform system C, and improves the DER by about 8\% absolute. This is remarkable because the proposed system is unsupervised and relatively computationally inexpensive when compared to LIUM (that utilizes i-vector based solution). We believe the better performance was achieved because CRSS-PLTL data contained shorter speaker turns, where the proposed system outperformed LIUM. Further analysis of DER across each audio channel revealed that the DER for individual channels varied between 22.48\% to 26.84\%, that suggests stable performance.
%-1mm}
\begin{table*}[!t]
\centering
\caption{{\it Comparison of proposed system and LIUM baseline using Diarization Error Rate (DER) and Speech Activity Detection (SAD) Equal Error Rate (EER).}}
\label{table_baseline}
%2mm}
\centerline{
%\begin{tabular}{*{3}{|c|}}
\begin{tabular}{*{3}{|c|c|c}}
\hline
System Used & DER (\%)& EER(\%)\\
\hline
LIUM (A) & 35.80 & 12.54\\
\hline
A + TO-Combo-SAD (B) & 34.20 & 8.67\\
\hline
B + Primary Speaker Identification (C) &  32.76& 8.67\\
\hline
Proposed System & \textbf{24.96} & \textbf{8.67} \\
\hline
%Total &  \\
%\hline
\end{tabular}
}
\end{table*}
%-2mm}
%%
\begin{table}[t]
\centering
\caption{{\it Showing performance of conversational-turn taking analysis using proposed speaker diarization system.}}
\label{table_turns}
%-2mm}
%\begin{tabular}{*{3}{|c|}}
\begin{tabular}{*{3}{|c|c|c}}
\hline
Member&Estimated Turns-taken& Error (\%)\\
\hline
Student 1 & 34& 5.56\\
\hline
Student 2 & 38& 6.45 \\
\hline
Student 3 & 27& 6.86 \\
\hline
Student 4 & 35& 7.32 \\
\hline
Student 5 & 39 & 4.88\\
\hline
Student 6 & 37& 6.45 \\
\hline
Leader & 37& \textbf{2.70} \\
\hline
\textbf{Mean} & \textbf{35.29} & \textbf{5.75} \\
\hline
\end{tabular}
%\label{table_turns}
\end{table}
\begin{figure*}[!t]
\centering
\includegraphics[width=450bp]{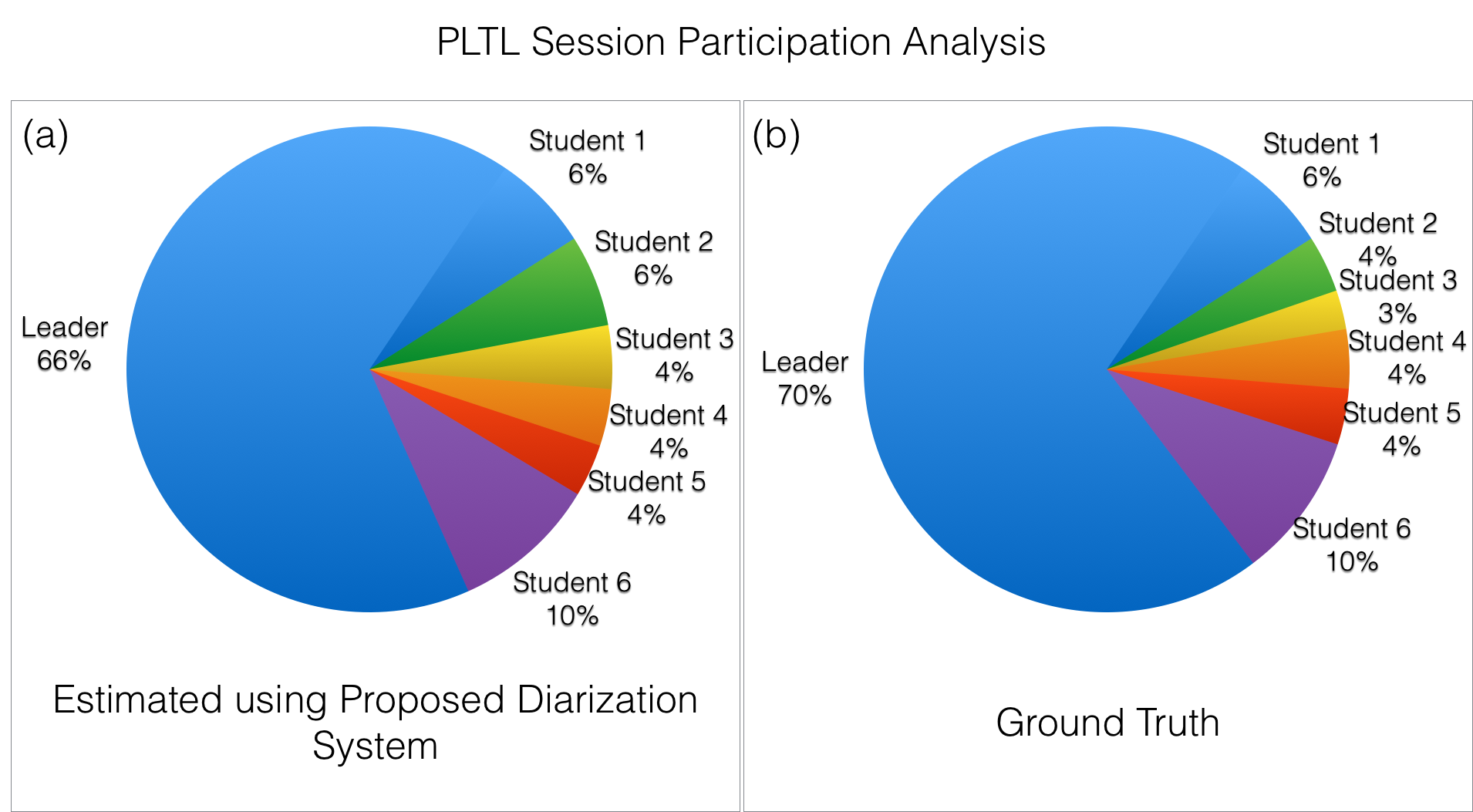}
\caption{{\it Automatic PLTL member participation analysis using proposed diarization system and comparison to analysis generated from ground-truth labels.}}
\label{fig_participation}
\end{figure*}
%-1mm}
Finally, we show two analyses using the proposed system. In the first analysis, the diarization output was used to count turns taken by each student and the team leader. The speaker turns could also be estimated from the ground-truth and this was used to determine accuracy of turn taking analysis. Table ~\ref{table_turns} shows turn taking estimation performance. It can be seen that the percentage error varies between 2.7\% and 7.32\%, that was interesting given that DER was about 24\% for this task. On average, each member took 35-to-36 turns in the 21-minute evaluation audio. Finally, we estimated how long each member spoke, by using the diarization output. In Fig.~\ref{fig_participation} (a), the proportional duration (that indicates proportional participation in conversation) is shown, and compared to a proportional participation pie chart generated using ground-truth in Fig.~\ref{fig_participation} (b). Comparison of the percentage participation numbers showed that the error were surprisingly low, and the analysis generated through proposed diarization method was rather accurate. For example, the leader occupies the conversation for almost two-thirds of the time, and students 6 and 3 contribute the most and least among students, respectively. In future work, encouraged by the results seen here, we wish to expand such analysis to the entire CRSS-PLTL corpus, and explore the ability to detect students at risk for subject material learning.
%-2mm}
\section{Conclusions}
%-1mm}
This study proposed an unsupervised speaker diarization system that used a new speaker change detection algorithm (termed unsupervised $G^{3}$ algorithm) and a new speaker clustering algorithm based on Hausdorff distance. A feature set for unsupervised $G^{3}$ algorithm that worked well for PLTL data had also been proposed. TO-Combo-SAD was used to separate speech from non-speech. The proposed diarization system was evaluated on a new corpora called CRSS-PLTL. The new corpora presents opportunity for speaker diarization research and its application in education research. In the experimental evaluations shown, the proposed diarization system significantly outperform the baseline LIUM diarization system. Finally, practical analysis using the proposed diarization system output was presented and discussed. The results and analyses presented are encouraging and motivate use of speech processing technology in studying practical problems in education research in particular, and human-to-human communication problems for small groups in general.
%  \section{Conclusions}
% {\color{blue}
% To be added }
% \section{Acknowledgements}
%%The ISCA Board would like to thank the organizing committees of the past INTERSPEECH conferences for their help and for kindly providing the template files.
%%\newpage
%\eightpt
%\nocite{*}
%\newpage
%\eightpt
\bibliographystyle{IEEEtran}
\bibliography{PLTL_diar}
  %\bibliography{mybib}
%  \begin{thebibliography}{9}
%    \bibitem[1]{Davis80-COP}
%      S.\ B.\ Davis and P.\ Mermelstein,
%      ``Comparison of parametric representation for monosyllabic word recognition in continuously spoken sentences,''
%  \textit{IEEE Transactions on Acoustics, Speech and Signal Processing}, vol.~28, no.~4, pp.~357--366, 1980.
% \bibitem[2]{Rabiner89-ATO}
% L.\ R.\ Rabiner,
% ``A tutorial on hidden Markov models and selected applications in speech recognition,''
% \textit{Proceedings of the IEEE}, vol.~77, no.~2, pp.~257-286, 1989.
% \bibitem[3]{Hastie09-TEO}
%   T.\ Hastie, R.\ Tibshirani, and J.\ Friedman,
%      \textit{The Elements of Statistical Learning -- Data Mining, Inference, and Prediction}.
%  New York: Springer, 2009.
%  \bibitem[4]{YourName16-XXX}
%  F.\ Lastname1, F.\ Lastname2, and F.\ Lastname3,
%Title of your INTERSPEECH 2016 publication,''
%in \textit{Interspeech 2016 -- 16\textsuperscript{th} Annual Conference of the International Speech Communication Association, September 8–12, San Francisco, California, USA, Proceedings, Proceedings}, 2016, pp.~100--104.
%  \end{thebibliography}
\end{document}